\documentclass[%
 aip,
 amsmath,amssymb,
 reprint,%
]{revtex4-1}

\usepackage{graphicx}
\usepackage{dcolumn}
\usepackage{bm}
\graphicspath{ {./figs/} }

\usepackage[utf8]{inputenc}
\usepackage[T1]{fontenc}
\usepackage{mathptmx}
\usepackage[version=4]{mhchem}
\usepackage{color}
\usepackage{url}

\newcommand{\kB}{{k_{\text{B}}}}
\newcommand{\boltzfac}{{\exp\left(\frac{\varepsilon}{\kB T}\right)}}
\newcommand{\subred}{{_{\text{red}}}}
\newcommand{\subox}{{_{\text{ox}}}}
\newcommand{\Ef}{{E_{\text{f}}}}

\begin{document}

\preprint{AIP/123-QED}

\title{Marcus-Hush-Chidsey Kinetics at Electrode-Electrolyte Interfaces}

\author{Rachel Kurchin}
\affiliation{%
Department of Mechanical Engineering, Carnegie Mellon University, Pittsburgh, Pennsylvania 15213
}
\affiliation{Department of Materials Science and Engineering, Carnegie Mellon University, Pittsburgh, Pennsylvania 15213}

\author{Venkatasubramanian Viswanathan}
\affiliation{%
Department of Mechanical Engineering, Carnegie Mellon University, Pittsburgh, Pennsylvania 15213
}
\affiliation{Department of Materials Science and Engineering, Carnegie Mellon University, Pittsburgh, Pennsylvania 15213}

\date{\today}

\begin{abstract}
Electrochemical kinetics at electrode-electrolyte interfaces limit performance of devices including fuel cells and batteries.  While the importance of moving beyond Butler-Volmer kinetics and incorporating the effect of electronic density of states of the electrode have been recognized, a unified framework that incorporates these aspects directly into electrochemical performance models is still lacking.  In this work, we explicitly account for the DFT-calculated density of states numerically in calculating electrochemical reaction rates for a variety of electrode-electrolyte interfaces. We first show the utility of this for two cases related to Li metal electrodeposition and stripping on a Li surface and a Cu surface (anode-free configuration).  The deviation in reaction rates is minor for cases with flat densities of states such as Li, but is significant for Cu due to nondispersive $d$-bands creating large variation.  Finally, we consider a semiconducting case of a solid-electrolyte interphase (SEI) consisting of LiF and \ce{Li2CO3} and note the importance of the Fermi level at the interface, pinned by the redox reaction occuring there.  We identify the asymmetry in reaction rates as a function of discharge/charge naturally within this approach. The analysis code used in this work is available open-source on Github.
\end{abstract}

\maketitle

\section{Introduction}
Reactions at electrode-electrolyte interfaces control the limits of operation of various electrochemical devices.  In the case of batteries, interfacial kinetics control the rate of reactions at the anode and cathode, which ultimately lead to limits on fast discharge and charge capability of modern Li-ion batteries.\cite{Kerman2017,Yu2018}  Fast charging is an important requirement for electric vehicles,\cite{Albertus2017} while fast discharge is necessary for electric vertical take-off and landing aircraft.\cite{fredericks2018performance} In particular, there is significant interest in modifying the reaction rates of electrodeposition and stripping of Li metal electrodes by modifying the nature of the so-called solid electrolyte interphase (SEI).\cite{Hatzell2020}

The typical kinetic rate law used for Li metal electrodes is Butler-Volmer.\cite{doyle1993modeling}  Recently, Boyle et al.\cite{boyle2020} highlighted the importance of moving beyond Butler-Volmer kinetics and incorporating effects of reorganization within the Marcus-Hush-Chidsey (MHC) formalism\cite{marcus1956theory,chidsey} within in the low overpoential limit.  In an accompanying paper for this Special Issue, we extend that analysis and show the importance of accounting for the full MHC kinetic rate behavior.\cite{sripad2020}  However, most treatments assume that the electrode density of states (DOS) is constant and independent of the energy states/overpotential.  

In this work, we incorporate the effect of the DOS at the electrode using density functional theory calculations directly and numerically show its importance for a variety of cases.  We begin by discussing the effect of the DOS for Li electrodeposition and stripping for Li metal electrodes.  Given the relatively flat nature of the DOS of Li metal, it affects the kinetic rates to a lesser extent.  Next, we discuss the case of metal electrodeposition on Cu.  Given the DOS associated with the more localized $d$-band states, there is significant deviation between the kinetic behavior when accounting for its variation compared to assuming a constant DOS.  This highlights that the electronic states of the metal are an important factor in determining high-rate performance on anode-free cells.\cite{Pande2019}  Finally, we discuss the case of an electrochemical redox reaction with the density of states determined by example SEI components, LiF and \ce{Li2CO3}.

We also provide both our full analysis code open-source on Github as well as a user-friendly interface for visualization of the effect of different model parameters on the results presented herein that readers can view online.

\section{Methods}
\subsection{Theoretical Approach}

The insight of Chidsey~\cite{chidsey}, namely, that it is critical to consider the occupation of electronic states when assessing rate constants within a Marcus-type theory, was important. Building on this insight, the rate expressions developed in MHC theory (we here adapt notation from Reference~\citenum{marcus_symmetry_1996}) are:

\begin{equation}
    k\subred \propto \int_{-\infty}^\infty|V(\varepsilon)|^2\exp\left(-\frac{(\lambda-e\eta+\varepsilon)^2}{4\lambda \kB T}\right)f_{\text{FD}}(\varepsilon)\mathrm{d}\varepsilon\,,
    \label{mhcred}
\end{equation}
\begin{equation}
    k\subox\propto\int_{-\infty}^\infty|V(\varepsilon)|^2\exp\left(-\frac{(\lambda+e\eta-\varepsilon)^2}{4\lambda \kB T}\right)\big(1-f_{\text{FD}}(\varepsilon)\big)\mathrm{d}\varepsilon\,,
    \label{mhcox}
\end{equation}
for the reductive and oxidative directions, respectively, where $V(\varepsilon)$ denotes a coupling constant between the (presumed localized) states on the electrolyte molecule and the (presumed plane-wave-like) states in the solid electrode. This coupling constant includes the density of states as well as an overlap integral term. $\lambda$ is the reorganization energy, $e$ the electronic charge, $\eta$ the applied overpotential, $\kB$ Boltzmann's constant, and $T$ the absolute temperature. All energies $\varepsilon$ are measured relative to the Fermi level $\Ef$ and $f_{\text{FD}}$ denotes the Fermi-Dirac distribution for occupation of states by electrons:

\begin{equation}
    f_{\text{FD}}(\varepsilon) = \frac{1}{1+\boltzfac}\,,
\end{equation}
and $(1-f_{\text{FD}})$ the distribution for holes.

In prior analyses, either explicitly or implicitly, the $|V(\varepsilon)|^2$ term is presumed to have weak energy dependence and brought outside the integral. This allows the simplification presented in Reference~\citenum{zeng2014}, taking advantage of the fact that the integral is over the whole real line and the Fermi-Dirac distributions for electrons and holes can be equivalently thought of as complements of each other, or equal but with the energy coordinate running in the opposite direction:
\begin{equation}
    k_{\text{ox/red}} \propto \int_{-\infty}^{\infty} \exp\left(-\frac{(\varepsilon-\lambda\pm e\eta)^2}{4\lambda\kB T}\right)\frac{\mathrm{d}\varepsilon}{1+\boltzfac}\,.
    \label{mhc_orig}
\end{equation}
However, this treatment neglects the electrode's DOS, which generally \textit{does} have strong variations with energy, especially over scales larger than a few $\kB T$. The most extreme example of this is in a semiconductor, where the value is identically zero at and near the Fermi level. This, as we will show, can lead to dramatically different rate constant predictions. However, even in the case of a transition metal (which often has large spikes in the DOS due to highly localized $d$-bands), this assumption can break down.

In our analysis, we consider adapted versions of Equations~\ref{mhcred} and~\ref{mhcox} where we explicitly account for the electrode DOS $\mathcal{D}(\varepsilon)$:

\begin{equation}
    k\subred \propto \int\mathcal{D}(\varepsilon)\exp\left(-\frac{(\lambda-e\eta+\varepsilon)^2}{4\lambda \kB T}\right)\frac{\boltzfac}{1+\boltzfac}\mathrm{d}\varepsilon\,,
    \label{kred}
\end{equation}
\begin{equation}
    k\subox\propto\int\mathcal{D}(\varepsilon)\exp\left(-\frac{(\lambda+e\eta-\varepsilon)^2}{4\lambda \kB T}\right)\frac{\mathrm{d}\varepsilon}{1+\boltzfac}\,,
    \label{kox}
\end{equation}

Because DOSes are not defined over an infinite range of energies, we must truncate the integrals in our calculations. In practice, this makes no discernible difference to the predicted rate constants, because the Gaussian terms fall off dramatically within a relatively small energy range from $\Ef$.

We note also that the dependence of the density of states (and to a lesser extent, the coupling constant that is also part of the $|V(\varepsilon)|^2$ term in Equations~\ref{mhcred} and~\ref{mhcox}) has been described before~\cite{Royea2006}, but is generally either assumed to be weak, or is modeled with simple functional forms that presume e.g. spherical Fermi surfaces or other approximations that may or may not be valid across the wide diversity of relevant surfaces. To our knowledge, this is the first analysis that explicitly accounts for the DFT-calculated DOS numerically in directly calculating rates for electrochemical reactions at electrode-electrolyte interfaces.

\subsection{Computational Approach}
In this work, we use the Julia programming language to calculate the integrals described above using adaptive Gauss-Kronrod quadrature~\cite{GK} as implemented in the QuadGK.jl package.

Density functional theory calculations for Li and Cu were conducted using the GPAW package~\cite{GPAW} through the Atomic Simulation Environment package.~\cite{ASE} Ion-electron interactions were treated using the Projector Augmented Wave approach.~\cite{PAW} For all calculations, a grid spacing of 0.16~\AA~and a 4 $\times$ 4 $\times$ 1 Monkhorst-Pack k-mesh were used.~\cite{MP} To improve self-consistent field convergence, Fermi smearing was applied to electron occupation with a width of 0.05 eV. All relaxations and analysis, unless otherwise specified, were conducted using the BEEF-vdW exchange-correlation functional. DOSes were calculated with a 100 meV Gaussian smearing width.

Calculations for LiF/\ce{Li2CO3} were performed in Quantum Espresso according to methods described in Reference~\citenum{Ahmad2020}.

\section{Results and Discussion}
\subsection{Li metal anode}
 
\begin{figure}
    \centering
    \includegraphics[width=0.95\columnwidth]{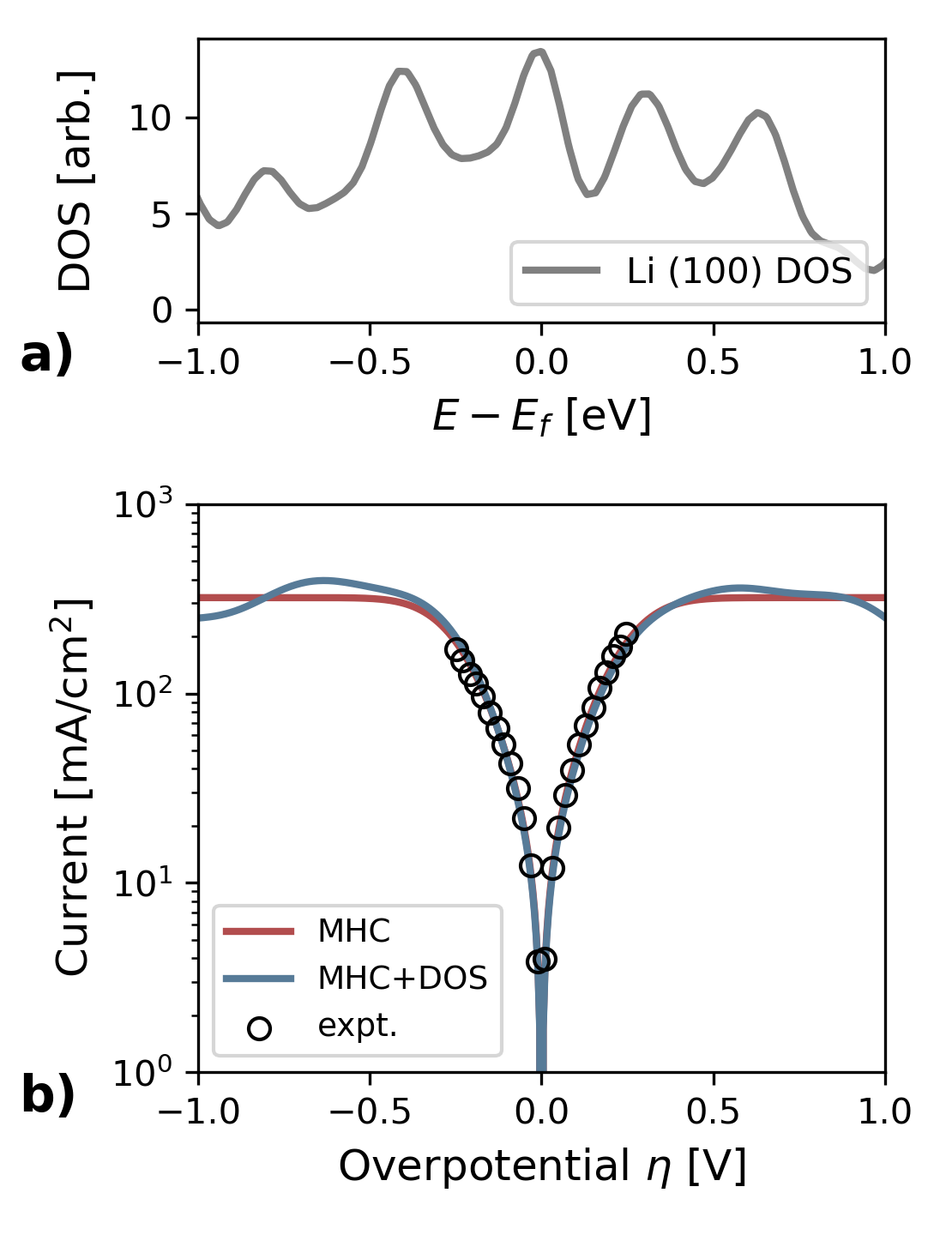}
    \caption{a) DFT-calculated DOS of Li (100) surface. b) Fitting results of the traditional MHC model and the revised MHC model with DOS integration to experimental data~\cite{boyle2020} of EC:DEC electrolyte at a Li surface. The resulting reorganization energies were 224 and 261 meV, respectively.}
    \label{fig:ecdec_li}
\end{figure}

 The first case is the redox processes occuring in a Li metal electrode.  The overal redox reaction is given by
 \begin{equation}
     \ce{Li+} + \ce{e-} \rightleftharpoons \ce{Li}\,,
 \end{equation}
where the forward reaction denotes electrodeposition (charging) and backward reaction denotes stripping (discharging).  The reaction rates are dependent on the organic electrolyte used\cite{boyle2020} and the analysis is presented for the case of an EC:DEC electrolyte. Figure~\ref{fig:ecdec_li} shows experimental data and best-fit results for each model (Results modeled with Equation~\ref{mhc_orig} are indicated in figure legends by ``MHC'' and from Equations~\ref{kred} and~\ref{kox} by ``MHC+DOS''). Note that over the experimental overpotential range of $\pm0.2$V the two models fit equally well. This is due to the fact that the DOS is relatively flat over this range. Over the full range of the plot, one can start to see variation between the models, but only modestly.
 
 This case illustrates why the importance of this more generalized analysis could have been missed previously -- at relatively modest overpotentials, many DOSes are ``close enough'' to flat that predictions of the traditional MHC model and the one we present here differ little. We will now turn to cases where the two models show larger discrepancies.

\subsection{Cu current collector}

\begin{figure}
    \centering
    \includegraphics[width=0.95\columnwidth]{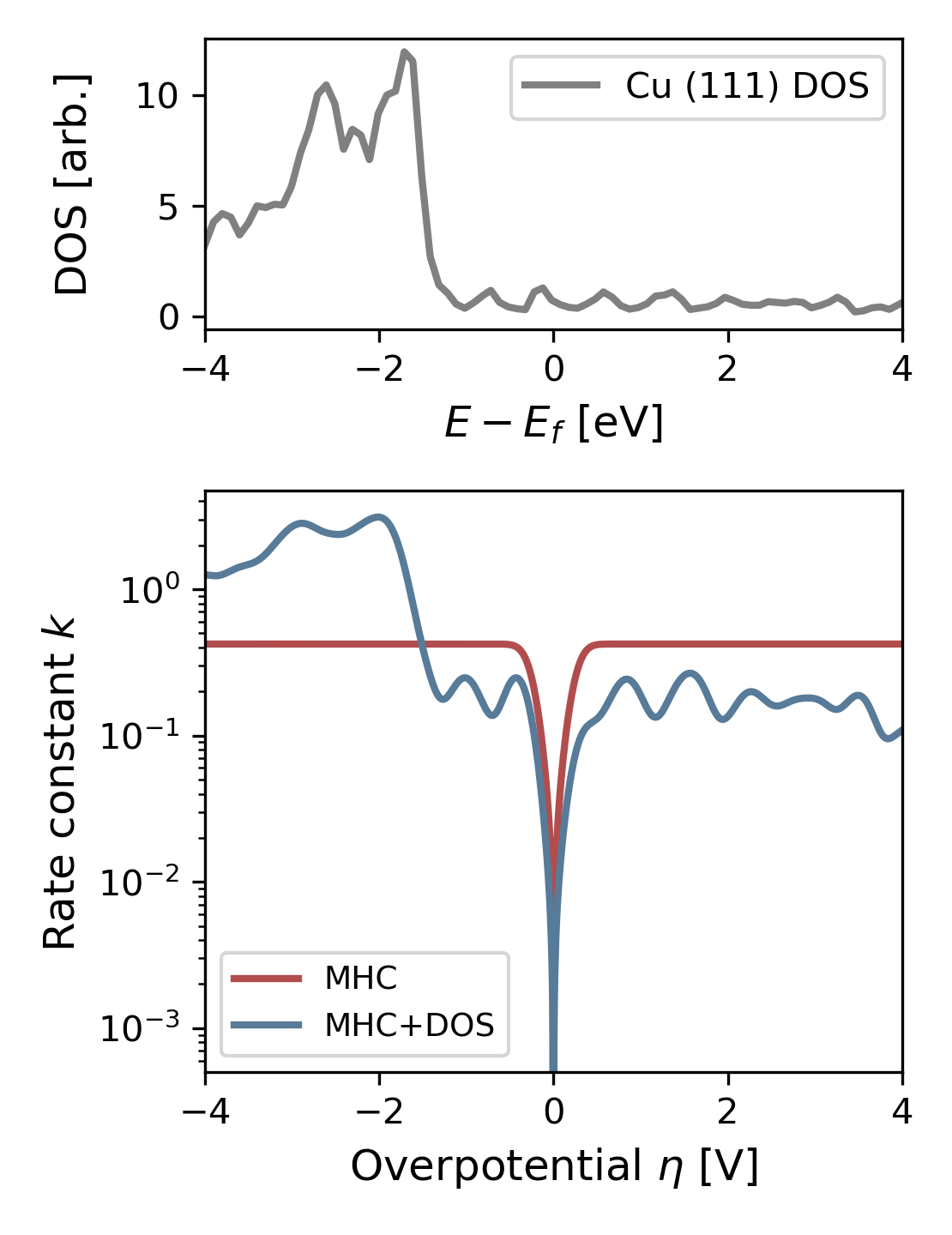}
    \caption{a) DFT-calculated DOS of Cu (111) surface. b) Predicted rate constants under the traditional and updated MHC models, using the same $\lambda$ values as Figure~\ref{fig:ecdec_li}.}
    \label{fig:cu}
\end{figure}

We next examine the case of depositing Li in an anode-free configuration on top of a Cu current collector.\cite{Pande2019}  This is given by the equation
 \begin{equation}
     \ce{Li+} + \ce{e-} \rightleftharpoons \ce{Li}^*\,,
 \end{equation}
where $*$ represents a site on a Cu surface.  In this case, the electronic states of the Cu surface become important. The DOS (a) and predicted rate constants (b) for the Cu(111) surface are shown in Figure~\ref{fig:cu}. We chose Cu(111) as it has the most desirable nucleation characteristics among the low Miller index surfaces,\cite{Pande2019} in addition to the lowest surface energy.\cite{Tran2016} Note that due to the high density of states in the Cu $d$-bands starting approximately 1.5 eV below the equilibrium $\Ef$, a jump of roughly an order of magnitude in the rate constant is observed at this overpotential. An overpotential of this magnitude is entirely feasible, particularly in the context of the fast charge/discharge rates currently targeted for many applications.

It is worth noting that the scale of variation in DOS (i.e. an order of magnitude) is approximately equal to the change in predicted rate constant. Indeed, one should expect this provided $\lambda$ is much smaller than the overpotential window of interest, since $\lambda$ essentially parametrizes a sliding Gaussian filter through which the DOS impacts $k$ via the denominator of the first exponential terms in Equations~\ref{kred} and~\ref{kox}.

\subsection{Effect of Solid-Electrolyte-Interphase (SEI)}
We will now consider the case where the Li electrode is covered with an inorganic solid electrolyte interphase due to spontaneous reaction with the organic electrolyte typically used in Li-metal batteries.  While the exact nature of the SEI remains an active area of research, it is widely acknowledged that some of the inorganic phases present include \ce{Li2CO3}, \ce{LiF}, \ce{Li2O}, etc.\cite{Li2017}  We consider a model SEI consisting of an interface between \ce{LiF} and \ce{Li2CO3}, building on the work of Pan et al.\cite{Pan2016}  Here, we consider a redox reaction, e.g. Li deposition, occurring at the SEI component.  Most SEI components are semiconducting or insulating by design and hence, the kinetics for this semiconductor-electrolyte interface are quite different.

\begin{figure}
    \centering
    \includegraphics[width=0.95\columnwidth]{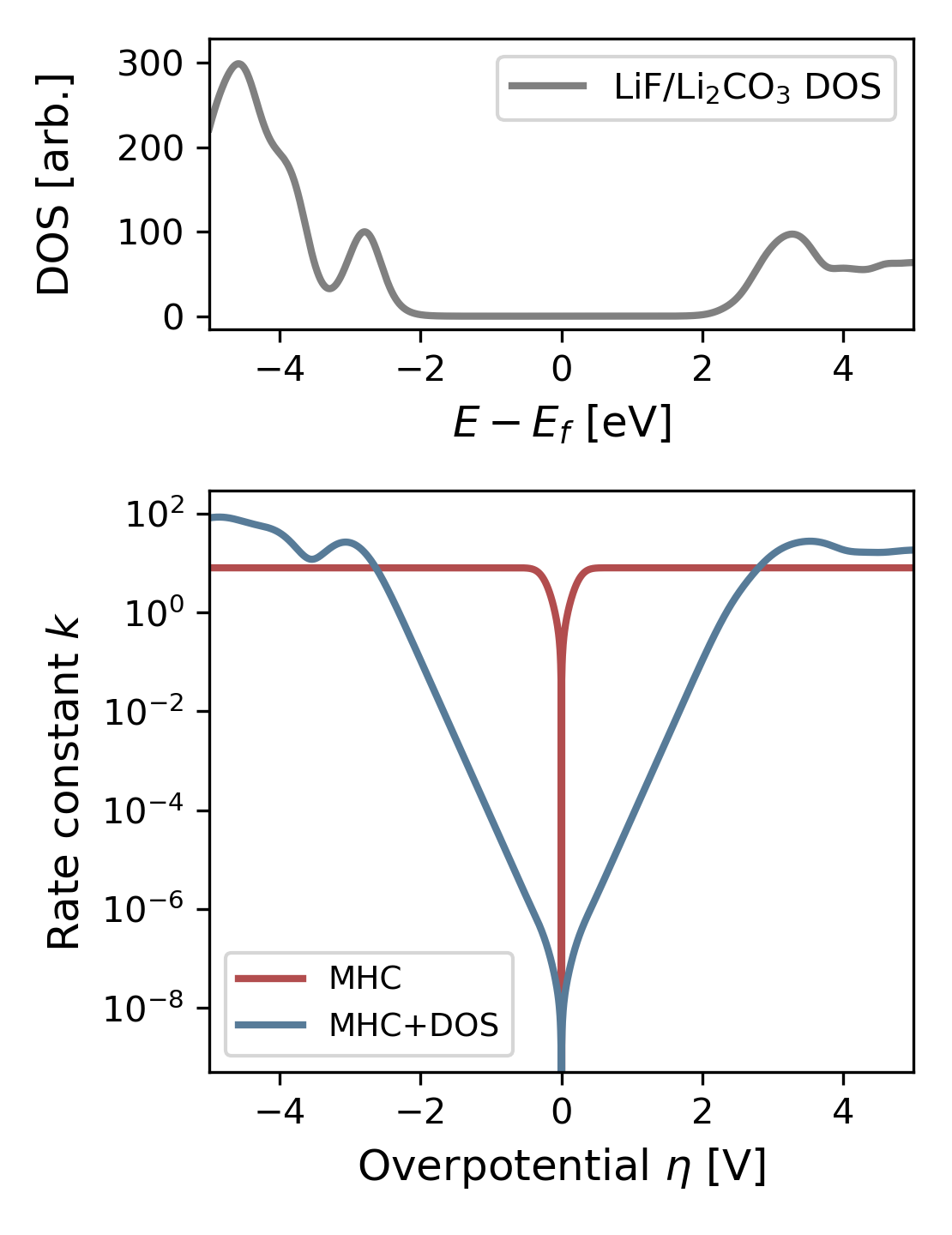}
    \caption{a) DFT-calculated DOS of LiF/Li$_2$CO$_3$ coherent interface generated by aligning the lattice vectors of LiF and \ce{Li2CO3} with angle $90^{\circ}$ between them and repeating the in-plane lattice vectors.\cite{Ahmad2020} b) Predicted rate constants under the traditional and updated MHC models, using the same $\lambda$ values as Figure~\ref{fig:ecdec_li}.}
    \label{fig:sei}
\end{figure}

Figure~\ref{fig:sei} shows the results of this modeling. Because the DOS is identically zero throughout the $\sim$4 eV bandgap, and the equilibrium $\Ef$ position of the pristine interface is at midgap, there is a discrepancy of as much as 8 orders of magnitude between the predictions of the two models, even at low overpotentials, indicating the criticality of considering the energy dependency of the DOS in this context. This further substantiates the remark above regarding the scale of variation of the DOS – because it goes to zero (over a window of many $\lambda$) in the bandgap, this scale tends to infinity, and the scale of the discrepancy in $k$ likewise diverges.


\subsection{Effects of model parameters}
Several parameters in the form of energy scales control the behavior of these models. Some of their behavior has already been alluded to above, and we examine these effects more closely here. First is the position of the equilibrium Fermi level $\Ef$. Thus far, we have assumed it to be at the position predicted by DFT for a pristine surface. However, many system variables can influence this energy. Bulk or surface defects can shift $\Ef$.  In an electrochemical system, the redox reaction sets the Fermi level at the electrode-electrolyte interface and is an independent variable.\cite{Viswanathan2014}

\begin{figure}
    \centering
    \includegraphics[width=0.95\columnwidth]{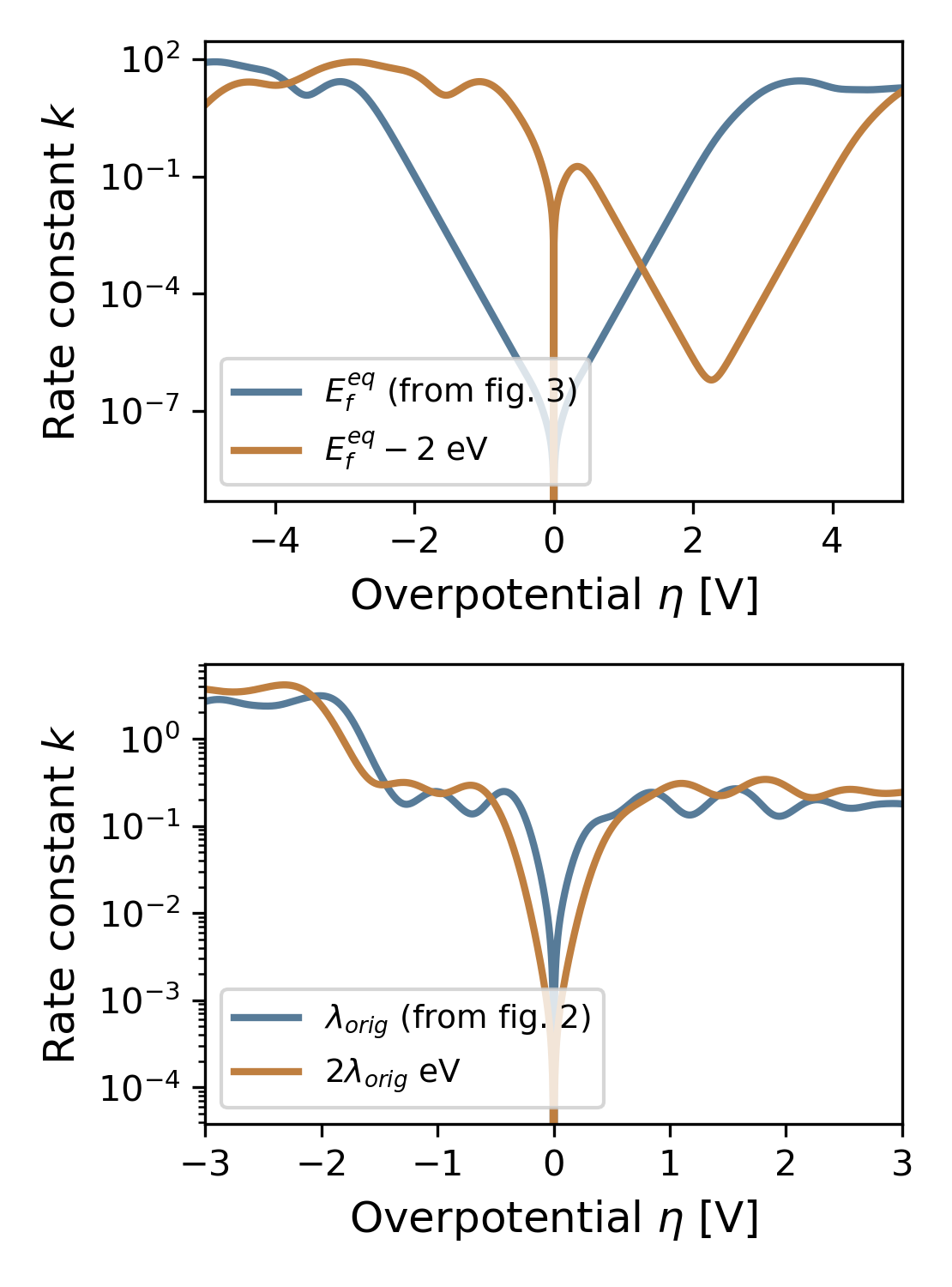}
    \caption{a) Rate constant for SEI with equilibrium (midgap) $E_f$ (reproduced from Figure~\ref{fig:sei} for reference), and with $E_f$ shifted 2 eV down, to near the valence band maximum. b) Rate constant for Cu (111) from Figure~\ref{fig:cu} compared to the same calculation but with the reorganization energy $\lambda$ doubled.}
    \label{fig:params}
\end{figure}

Figure~\ref{fig:params}a shows the effect of shifting $\Ef$ on the SEI case. In particular, we presume an alternative redox couple (or equivalently, a particular voltage on the Li/Li$^+$ scale) is chosen such that the equilibrium $\Ef$ is near the valence band maximum of the SEI.  In this case, the behavior at low overpotentials exhibits a marked asymmetry: at negative $\eta$ (pushing further into the valence bands), $k$ rises sharply, while at positive $\eta$ (pushing back into the bandgap), it drops exponentially until reaching midgap, whereupon it rises again as the conduction band minimum begins to have an impact. Such an asymmetry depending on the redox potential has been recognized\cite{Luntz2013} and observed\cite{Knudsen2015} in Li-\ce{O2} batteries where discharge product, \ce{Li2O2}, is a wide-band gap insulator similar to the LiF/\ce{Li2CO3} interface considered here. 

Figure~\ref{fig:params}b shows the effect of varying the reorganization energy $\lambda$. As mentioned previously, the mathematical effect of $\lambda$ is primarily to act as a Gaussian smoothing filter on the density of states. However, an increased $\lambda$ also increases the offset between the filters in the positive and negative direction, with the overall result being a slower varying and slightly outward-shifted $k$ vs. $\eta$ curve.

To better build intuition for the impact of these parameters, we have built an interactive visualization which can be viewed at \url{https://nbviewer.jupyter.org/github/aced-differentiate/MHC_DOS/blob/master/dataviz.ipynb}. It features the DOSes for all materials considered in this work, and allows the viewer to modify $\Ef$, $\lambda$, and the temperature, and see in real time how rate constant predictions change.

\section{Conclusions}
In this work, we explore the role of the electronic density of states at the electrode using density functional theory calculations and incorporate that numerically into a generalized Marcus-Hush-Chidsey kinetics formalism.  We discuss three cases, electrodeposition on (i) Li surface and (ii) Cu surface, and (iii) redox reactions at an SEI component chosen here to be LiF/\ce{Li2CO3} interface.  For the case of the Li surface, due to a relatively flat density of states, we find minor deviations from the standard MHC rate law.  In the case of a Cu surface, mimicking an anode-free Li metal battery, the $d$-bands of Cu can significantly modify the rate constants, especially at high rates of charge and discharge.  Finally, we show the drastic changes in reaction rates for a semiconducting interface and note the importance of the location of the Fermi level, which is pinned by the redox reactions occurring at that interface.  We show strong asymmetric behavior for the SEI case, similar to that noted for \ce{Li2O2} in Li-\ce{O2} batteries. We have also made all analysis code available open-source so that the community can take advantage of these models.

\section{Acknowledgement}
R.K. and V.V. gratefully acknowledge support from the Advanced Research Projects Agency Energy (ARPA-E) under Grant DE-AR0001211. This research was also supported by the Carnegie Mellon University Manufacturing Futures Initiative, made possible by the Richard King Mellon Foundation.  We thank Yumin Zhang, Shang Zhu and Zeeshan Ahmad for sharing structures used in this work.

\section{Data Availability}
The data and code that support the findings of this study are openly available on Github at \url{https://github.com/aced-differentiate/MHC_DOS}.

\bibliography{aipsamp}

\end{document}